\documentclass[aps,prstper,twocolumn,showpacs,letterpaper,longbibliography,nofootinbib,floatfix]{revtex4-2}

\usepackage{graphicx}
\usepackage{dcolumn}
\usepackage{bm}
\usepackage{hyperref}
\usepackage{multirow}
\usepackage{xcolor}
\usepackage{amsmath}
\usepackage{amssymb}

\usepackage{mathtools}

\DeclarePairedDelimiter\ket{\lvert}{\rangle}

\begin{document}

\preprint{APS/123-QED}

\title{Motivating reflection in problem solving: homework corrections in upper-division physics courses}

\author{Molly Griston} \email{molly.griston@colorado.edu} \author{Bethany R. Wilcox}
\affiliation{Department of Physics, University of Colorado, Boulder, Colorado 80309, USA}

\date{\today}

\begin{abstract}
Despite the recognition that reflection is an essential part of problem solving, it is often not emphasized in upper-division physics courses. In this paper, we discuss homework corrections (HWCs) as a pedagogical tool to motivate reflection on homework assignments. We focus on gaining a qualitative understanding of how students may engage with the process of homework corrections, considering potential impacts on content understanding, metacognitive skills, and affect. To do so, we present three comparative case studies to elucidate different aspects of student engagement with HWCs; we also present descriptive statistics of participation in HWCs. We find that HWCs can be a useful pedagogical tool, but more scaffolding may be necessary to support students in identifying and learning from their mistakes, even in upper-division courses. 


\end{abstract}

\maketitle


\section{Introduction}
An essential element of physics education is helping students become expert learners and problem solvers. To achieve this goal, both problem solving and metacognitive skills should be explicitly taught and practiced in physics courses \cite{zepeda2015direct, avargil2018students}. Significant work has been devoted to understanding both the problem solving process itself and how to teach students to become proficient problem solvers (e.g., \cite{reif1982knowledge, tuminaro2007elements, docktor2015conceptual, leak2017examining, burkholder2020template}). More broadly, efforts have been made to intentionally target students' metacognitive abilities in physics course design (e.g., \cite{mota2019homework, reinhard2022assessing}). A key theme of both problem solving and metacognition is reflection, which occurs during and after problem solving. In most upper-division physics courses, homework assignments are the primary place for students to practice problem solving. However, turning in the assignment often signifies the end of the problem solving process, and there is no formal structure in place for reflection after this step. Instructors typically post homework solutions, in varying levels of detail, and expect that students will engage with these solutions and use them to reflect on their own work. However, there is limited evidence to suggest students are actually engaging in this process and if they are, how effectively they are doing so \cite{mason_advanced_2010}. 

Here, there exists a natural opportunity to develop students' reflective skills. By engaging in scaffolded reflection on their homework assignments, students can reflect on both their performance on specific problems and their overall content understanding and engagement in the course. To motivate students to engage in this reflective process, we consider the implementation of homework corrections (HWCs), which we broadly define as any opportunity for students to make corrections on their assignments to improve their grade. 

HWCs can be viewed through the lens of self-assessment, in that they require comparing one's work with an external standard \cite{zhang_integrated_2023, andrade2019critical}. Even when the end product is graded, the opportunity for students to self-correct on their homework assignments is inherently formative. Significant work has demonstrated the value of formative self-assessment, showing that self-assessment can help students learn from their mistakes and support self-regulated learning \cite{andrade2019critical}. Within PER, researchers have studied the effects of self-diagnosis activities on subsequent performance, finding that both introductory and upper-division students can benefit from scaffolding to help them learn from their mistakes \cite{yerushalmi_physics_2007-1, singh_physics_2007, cohen_identifying_2008, yerushalmi_effect_2008, yerushalmi2009self, mason2009self, yerushalmi_what_2012-1, safadi_students_2013, brown_improving_2016, mason_learning_2016, safadi_designing_2017, mason_reworking_2022, mason_improving_2023}. Recent work specifically advocates for providing upper-division students explicit incentive to correct their mistakes \cite{brown_improving_2016, mason_improving_2023}. However, additional work is needed to gain insight into different implementation methods, the mechanisms involved, and how this activity fits into the larger picture of student engagement in physics courses. 


 In this paper, we conduct an exploratory analysis of how students are engaging with HWCs in an upper-divison physics course. We will introduce two specific methods of implementation, identify ways that students are engaging with HWCs and how these are related to motivations for and expectations around implementing HWCs, and discuss factors for instructors to consider when deciding if and how to implement HWCs. This analysis is intended to provide insight into students' thoughts and processes related to reflection and learning from errors, specifically in the context of upper-division physics courses, to inform future implementation efforts. We focus on presenting a holistic view of what engagement with HWCs can look like for some students; as such, we are prioritizing depth of analysis, rather than breadth, and are not intending to make any generalizable statements about the impact of HWCs on all students. 

In Section \ref{sec:background}, we will introduce relevant background information, focusing on reflection in problem solving, self-assessment, and error climates in the classroom (i.e., the extent to which the environment supports or inhibits learning from errors \cite{STEUER2013196}). We will then discuss our prior work related to student and instructor motivations and concerns regarding HWCs in Section \ref{sec:priorwork} \cite{griston_homework_2024}. In Section \ref{sec:methods}, we will describe the relevant context, as well as our methods used for data collection and analysis. This paper includes several different complementary sources of data, including coursework, surveys, and interviews (with both semi-structured and think-aloud sections). In Section \ref{sec:results}, we will first present descriptive statistics, which are intended to give context to how many students are engaging with HWCs, as well as which students are doing HWCs. Then, we will present a multiple case study analysis of three students in the same course; this will include within-case and cross-case analyses. The within-case analysis is intended to provide a detailed perspective of what engagement with HWCs can look like for different students. The cross-case analysis then addresses trends across the cases, with specific attention paid to ways in which engagement does or does not align with instructors' intentions and motivations for implementing HWCs, as well as aspects of engagement outside the original HWCs framework. Our analysis is specifically guided by the following questions:
\begin{itemize}
    \item[RQ1:] In what ways and for what reasons are students engaging with HWCs? How (if at all) are their stated motivations and goals related to their behavior?
    \item[RQ2:] Are students engaging with HWCs in ways that are not within the original HWCs framework (as established by the instructor) but may be productive? How does student engagement align with instructor motivations for implementing HWCs? 
    \item[RQ3:] Are there differences in how different students engage with HWCs?
    \item[RQ4:] Where (if at all) are students struggling when engaging with HWCs? What insight can this provide regarding students' reflective processes?
    \item[RQ5:] In what ways (if at all) might choices in the implementation of HWCs impact the nature and frequency of student engagement?
\end{itemize}
Finally, in Sections \ref{sec:discussion} and \ref{sec:implications} we will provide a discussion of the results and instructional implications. 

\section{Background}\label{sec:background}

This work intersects with several related areas that have been discussed in the literature, including reflection as it pertains to problem solving, self-assessment literature as it pertains to reflection, and classroom error climates as they relate to students' affective experience when reflecting on their work and learning from mistakes. In this section, we will review literature relevant for each of these areas, particularly as they relate to our research questions. 


\subsection{Reflection in physics problem solving}

Within PER and STEM education more broadly, problem solving has been studied extensively. Researchers have consistently identified that reflection is an important part of the problem solving process. Reif and Heller's early work to develop a model of problem solving in physics includes the ``Exploration of Decisions" throughout the process, in which one identifies alternate choices, compares their corresponding utilities, and revises unsatisfactory choices as necessary. Once having reached a solution, they argue the problem solver should assess their solution and revise as necessary until it meets the following criteria: clear interpretation, completeness, internal and external consistency, and optimality \cite{reif1982knowledge}. More recently, Salehi presented a problem solving framework that splits problem solving practices into experimental practices and reflective practices, in which experimental practices are done to solve the problem and reflective practices are done to make intentional and informed decisions about the experimental practices \cite{salehi_improving_2018}. Studying science and engineering experts, Price et al. found that reflection decisions occur throughout the problem solving process and are ``critical for success" \cite{price2021detailed}. While the details have differed between specific problem solving frameworks, the importance of reflection during problem solving has remained consistent.

Due to the importance of reflection in problem solving, efforts have been made to explicitly teach reflection skills in the physics classroom. For example, Burkholder et al. developed a problem solving framework that asks students to engage in a series of ``Answer Checking" steps \cite{burkholder2020template}. Other work has looked at the impact of teaching students specific evaluation skills (e.g., unit analysis), finding that such instruction can have a significant impact on students' problem solving skills \cite{warren2010impact}. These efforts, however, have focused on students' abilities to reflect without any external evaluation or point of comparison. While such skills are important, there are also many instances, both in and out of the classroom, when students are provided with problem solutions, with varying degrees of detail, that can be used to reflect on their own problem solving process. Thus, it is also necessary to develop students' reflective skills in this context. Additionally, engaging in reflective practices with reference material can help students develop the skills required to engage in the aforementioned reflective steps without such materials. 

\subsection{Self-assessment}

To contextualize this reflective process, we turn to the self-assessment literature, as self-assessment is often structured to include comparison of one's own work to an external source, such as a provided solution. Within self-assessment, it is important to provide students the opportunity to adjust and correct their own work, rather than merely assess it \cite{andrade2019critical}. Formative self-assessment can present an opportunity for students to learn from their errors by generating internal feedback regarding content understanding and the learning process \cite{zhang_integrated_2023, yan2017cyclical, butler1995feedback, nicol2021power}. Researchers argue that to generate high-quality internal feedback and learn from errors, students must engage in self-explaining, which refers to ``the activity of generating explanations to oneself, usually in the context of learning from an expository text." Within self-explaining, the mechanism proposed to be responsible for promoting learning is self-repair, in which students repair their mental models after identifying conflicts \cite{butler1995feedback, zhang_integrated_2023, chi_self-explaining_2000}. Therefore, when analyzing student engagement with HWCs, it is important to consider if and when students are self-explaining. This framework, however, only addresses one possible element of the self-assessment process, and there is minimal research analyzing the broader process of self-assessment and the mechanisms involved \cite{andrade2019critical}. Models that have been developed (e.g., \cite{yan2017cyclical, liu2022exploring}) are fairly high-level and focus on notably different implementations of self-assessment, which motivates our exploratory approach. 




 Within PER, researchers have used the self-explaining literature along with the cognitive apprenticeship model (an approach to providing scaffolding as students are enculturated in expert practices) \cite{dennen2008cognitive} to frame the development of self-diagnosis (SD) activities designed to target students' reflective skills. SD activities have been implemented in K-12 environments \cite{yerushalmi_physics_2007-1, safadi_students_2013, brown_improving_2016, safadi_designing_2017}, introductory physics courses \cite{yerushalmi_physics_2007-1, singh_physics_2007, cohen_identifying_2008, yerushalmi_effect_2008, yerushalmi2009self, mason2009self, yerushalmi_what_2012-1, mason_learning_2016}, and upper-division physics courses \cite{brown_improving_2016, mason_reworking_2022, mason_improving_2023}. These studies suggest that despite the beliefs of some professors that advanced students have learned to self-monitor and will therefore learn from their mistakes, even advanced students do not automatically do so \cite{mason_learning_2016}. Rather, instructors can motivate this behavior by rewarding students for correcting their mistakes (e.g., by giving points back on an exam for corrected work) \cite{brown_improving_2016, mason_reworking_2022, mason_improving_2023}. However, evidence shows that self-diagnosis activities may not lead to self-repair processes \cite{safadi_students_2013} and that intentional scaffolding is necessary, even for upper-division students \cite{mason_improving_2023}.

While these studies provide motivation for this work and have relevant takeaways, the SD activities have been designed to accompany traditionally more summative assessments (i.e., quizzes and exams). As students have different attitudes and stress levels regarding homework versus quizzes and exams, it is reasonable to expect that their engagement in SD activities would be different as well. On homework assignments, students have functionally unlimited resources available to them, whereas on quizzes and exams, their resources are much more limited (if available at all). It is a seemingly different activity for a student to reflect on a problem for which they had plenty of time and resources than on one where they did not. Additionally, SD activities for exams can only be done a few times in a semester, whereas homework assignments are typically weekly. We may expect different trends when students are engaging in scaffolded reflective practices more consistently, as opposed to a single time. It may alternately be the case that students are less motivated to complete corrections for homework assignments because an individual homework assignment makes up much less of their final grade than an exam. These differences motivate the need to conduct an exploratory analysis of student engagement with HWCs and compare these results to existing work. 

\subsection{Error climates}

As established in the previous section, errors can present an opportunity for students to learn; in fact, generating errors is beneficial for learning when accompanied by corrective feedback \cite{zhang_integrated_2023, metcalfe2017learning}. However, this opportunity is often limited by student and instructor beliefs and attitudes related to making errors. When surveyed, most instructors reported believing that errors are a normal part of the learning process and did not believe that students should try to avoid errors during learning; however, the majority of students reported avoiding making errors and associated errors with negative emotional consequences \cite{pan_learning_2020}. In the classroom, students are often given implicit messaging that errors are to be avoided, through the course structure. For example, giving entirely summative assessments with no opportunities for revision communicates to students that they must avoid errors. 

A course's messaging about making and learning from errors constitutes its error climate. Having a positive error climate can enhance learning; specifically, both the emotional and cognitive aspects of the error climate can impact students' abilities to learn from errors \cite{steuer_constructive_nodate}. We hypothesize that implementing HWCs can positively impact a course's error climate. By formalizing the reflection process, instructors are communicating to students that making errors is a valuable part of learning and an expected step in the problem solving process. Additionally, HWCs may reduce the pressure students feel when initially submitting a homework assignment and therefore decrease their motivation to engage in behaviors such as copying from peers or using external resources (e.g., online solutions or generative AI). The impact of a courses' error climate on students' abilities to learn from mistakes emphasizes the importance of studying HWCs from a holistic perspective, with consideration for student engagement in the course as a whole, rather than in isolation.

\section{Prior Work}\label{sec:priorwork}

In our previous work, we focused on identifying instructor motivations and concerns regarding implementing HWCs, as well as student motivations for engaging with them \cite{griston_homework_2024}. Identifying instructor motivations and concerns provides necessary insight when analyzing student engagement and discussing instructional implications. For example, instructors communicated concern about certain student behaviors, such as ``gaming the system," and we can address these concerns in our discussion of the results. Identifying student motivations helped guide the structure of the current analysis and will allow us to interpret our case study results in the context of the broader motivation data.

\subsection{Student motivations and concerns regarding HWCs}
To elucidate student motivations for engaging with HWCs, we first surveyed students over the course of one semester. For the two different weeks students were surveyed, we found that all students who completed HWCs reported they were at least partially motivated by improving their grade, but a notable portion of students also reported being motivated by improving their understanding of the content and learning from their mistakes. Of the students who did not complete HWCs, a plurality reported that they had intended to but forgot; this was closely followed by limitations on time. However, multiple students also reported not completing HWCs, at least in part because they thought they would be happy with their grade and/or already understood the content well enough. 

These surveys were supplemented by interviews of students in the surveyed population. In these interviews, we found that students are balancing the time required to complete HWCs, their grade on the homework assignment and/or course, and  their understanding of the content. This initial work was limited in that we only analyzed students' reported motivations and could not contextualize their responses with other aspects of their engagement with HWCs or the course more broadly. One of the goals of the analysis in this paper is to provide a more comprehensive portrayal of how students' reported motivations can interact with other aspects of their engagement in the course. 

\subsection{Instructor motivations and concerns regarding HWCs}

It is important to understand the motivations and concerns of instructors, both for making HWCs an effective pedagogical tool and interpreting and contextualizing students' experiences. In our previous work, we found that instructors were broadly motivated by wanting students to review the posted solutions and reflect on their own work. Some instructors spoke specifically of students learning the content on the assignment, while others discussed more metacognitive components, such as wanting students to reflect on the learning process. Instructors also explicitly discussed wanting to make homework more formative, as opposed to summative, as well as hoping to boost student morale and reduce the fear of making mistakes. Instructors were concerned, however, about the logistics involved in HWCs, wanting to avoid an increased grading load. Additionally, they noted that grading assignments with HWCs can require more nuanced grading, which may be more difficult for graduate teaching assistants. Some instructors were concerned that allowing students to do HWCs would mean the students could put minimal effort into the original assignment and ``game the system." Others were worried about inflating homework grades and making the homework portion of the grade ``less meaningful." However, these concerns were not universal, and several instructors explicitly voiced that they were not worried about inflating homework grades or students gaming the system. One issue that did not come up during instructor interviews but has arisen during informal conversation and is worth noting is that the implementation of HWCs, as described in this paper, essentially eliminates the possibility of students submitting late work. This concern, along with the others described in this section, will be discussed in greater detail when considering instructional implications in Section \ref{sec:implications}.

\begin{figure}[b]
\includegraphics[width=0.5\textwidth]{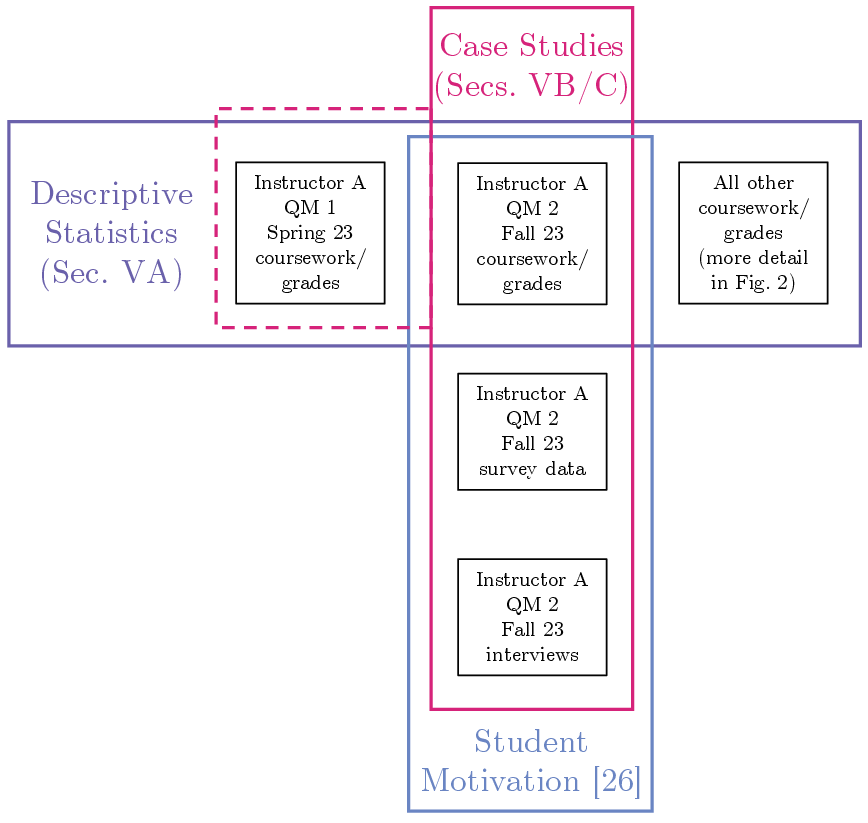}
\caption{\label{fig:datafig}Diagram of the different data sources used for analysis. The dashed line indicates that the data source was referenced but not explicitly part of the analysis. \textbf{Student Motivation} refers to our previous work, discussed in detail in \cite{griston_homework_2024}. \textbf{Descriptive Statistics} are discussed in Section \ref{sec:stats} and include coursework data from all available courses. The \textbf{Case Studies} are presented in Sections \ref{sec:within} and \ref{sec:between} and include all data collected in Instructor A's QM2 course (coursework/grades, surveys, and interviews) and reference Instructor A's QM1 course.}
\end{figure}

\section{Context and Methods}\label{sec:methods}

In this section, we will describe the relevant context and the methods used for data collection and analysis. Due to both the availability of data and goals of the analysis, this paper includes several overlapping sources of data. The data sources and corresponding analysis sections are shown in Fig. \ref{fig:datafig}. 

\subsection{Department, instructor, and course context}\label{sec:course}

This research was conducted in the Physics Department at the University of Colorado Boulder, which is a predominantly white, large research institution. The Physics Department awards approximately 100 undergraduate degrees annually. Further, the department has a significant history of educational reform at the undergraduate level, and many instructors use research-based practices in their courses (e.g., \cite{pollock2008sustaining, chasteen2015educational, pollock2023adaptable}). 

Data from the courses of two different instructors are included in this analysis. One of the instructors, Instructor A, is the second author of this paper (BRW). Both instructors are active in the PER community and have received teaching awards. Their courses are very similar with regard to structure and materials but are not identical; differences will be discussed where relevant. 

In Section \ref{sec:stats}, we reference a total of twelve courses, six of them taught by Instructor A and six of them taught by Instructor B.\footnote{One of the courses (Classical Mechanics and Math Methods 1) was co-taught by Instructor B and another instructor.} These twelve courses include four different subjects: Classical Mechanics and Mathematical Methods 1 (CM1), Thermodynamics and Statistical Mechanics (SM), Quantum Mechanics 1 (QM1), and Quantum Mechanics 2 (QM2). CM1 and QM1 are both the first course of a two-semester sequence and are required of all physics majors. SM is a single semester course and required of most majors. QM2 is the second course of a two-semester sequence and required of some majors. We also note that some of these courses took place in a remote or hybrid format during the COVID-19 pandemic. In all of these courses, homework made up between 30 and 40\% of the final grade.

\subsection{HWC implementation and logistics}

In Instructor A's courses, HWCs were implemented with the following structure. On Thursday nights, students turned in their homework assignments. Full, worked-out solutions were posted Friday morning. If students wanted to complete HWCs, they were due on Saturday at noon; however, the students were informed that they would receive credit for their HWCs as long as they were completed before the grader started grading which nearly always meant that corrections submitted by sometime on Sunday would get full consideration. All homeworks were then graded by a graduate TA. Students could receive full credit for appropriately corrected problems. 

For the corrections themselves, students were asked to use the Canvas commenting tool on their original submission to make a reflective statement about what they did wrong and write a description of what the correct approach would be; however, students did not have to fully rework the problem. They could only complete HWCs on problems they had reasonably attempted on the initial submission.\footnote{Students were not given strict guidance about what constituted a ``reasonably attempted" problem.} The students were given these instructions on the first day of class, as well as on Canvas. The following were provided as examples of patterns for comments that would get credit:

\begin{itemize}
    \item ``Even though I got the right answer, my justification was not fully correct because [...]."
    \item ``My argument was flawed because [...]."
    \item ``The correct answer involves recognizing that [...] happens because [...]."
\end{itemize}

The following were provided as examples of comments that would not get credit:

\begin{itemize}
    \item ``The solution said [...] should have been the correct answer."
    \item ``I got the same answer as the solution, I just did it in a less robust way."
    \item ``I did this part wrong, it should have been [...]."
\end{itemize}

In Instructor A's course, late homework assignments were not accepted, due to the implementation of HWCs. However, every student's lowest homework grade was automatically dropped and the instructor told the students that they could request a second dropped score in the case of extenuating circumstances. 

The structure of HWCs in Instructor B's courses was similar to that in Instructor A's course but varied in notable ways. Homework solutions were posted automatically at the deadline. Then, students would write their corrections directly on their initial assignment (e.g., on the physical paper in a different colored pen) and resubmit the assignment. For their corrections, students were asked to provide a ``thoughtful discussion" of the mistakes they made, why they made them, how they would avoid them in the future, and how to arrive at the correct solution (in detail). Between the two instructors, there were two primary differences in implementation. First, the format: Instructor A's students used Canvas's commenting tool to make comments on their submissions, and Instructor B's students resubmitted their assignments with annotations written in a different colored pen on the original assignment (or the digital analog for those who submitted typed assignments). Second, the content of the corrections: Instructor A did not require students to fully rework the problem, while Instructor B did. 

\subsection{Primary case studies}

Here, we describe the data collection relevant to the primary case studies, represented by the pink box in Fig. \ref{fig:datafig}. We then discuss the selection of the case studies and the methods of analysis. 

\subsubsection{Student surveys and interviews}\label{sec:studentsurveys}

During Instructor A's QM2 course, the students were surveyed four times as part of the course, on their weekly preflight quizzes (short, weekly quizzes graded for completion). The students were surveyed during the first week of the semester (S1), after their second homework assignment (S2), after the homework assignment following their first exam (S3), and during the last week of the semester (S4). 

On S1, the students were asked about their experience with HWCs in a previous semester:

\begin{quote}
    \emph{S1: In your QM1 course, how often did you complete HWCs? On weeks that you did complete the HWCs, what was your motivation? On weeks that you did not complete the HWCs, why not? Did you find the HWCs valuable? Why or why not?}
\end{quote}

Student responses to these open-ended questions about motivation were coded to generate ``select-all" questions administered on S2 and S3:
\begin{quote}
    
\emph{S2/S3: What motivated your decision to complete homework corrections (or not)?  Select all that apply
\begin{itemize}
    \item[$\square$] I wanted to improve my grade
    \item[$\square$] I wanted to better understand the content
    \item[$\square$] I wanted to learn from my mistakes
    \item[$\square$] I didn't have enough time
    \item[$\square$] I thought I would be happy with my original grade
    \item[$\square$] I already understood the content well enough
    \item[$\square$] I had intended to but forgot
    \item[$\square$] \textbf{(Only on S3)} My confidence about the first exam 
    \item[$\square$] Other (feel free to elaborate in the final text box)
\end{itemize}}
\end{quote}

S4 was given in the last week of the semester, and students were asked:

\begin{quote}
    \emph{If you were an instructor for an upper-division physics course, would you offer your students the chance to complete HWCs? If so, why and how would you structure it? If not, why not?}
\end{quote}

In addition to the surveys, nine students in Instructor A's QM2 course were interviewed during Week 11 of the semester. The interview solicitation was sent to all students in the course, and students were offered monetary compensation (\$20) for their participation. They were also informed that their instructor would not know whether or not they had chosen to participate and their participation would have no impact on their grade in the course. 

In the interviews, students were first asked to do their HWCs for the week in a think-aloud format. Since students were completing HWCs on their computers, both audio and screen recordings were collected. During this section, the interviewer only prompted students to remind them to verbalize their thoughts. Once students were done with their HWCs, they were asked questions, in a semi-structured format, regarding their overall experience with HWCs. The interviews lasted between approximately 15 minutes and an hour. All of the interviews were conducted by the first author.\footnote{We note that the first author was a grader for QM1 the previous semester, which all of the interview participants were enrolled in; however, she did not have any in-person interactions with them in that role, and she had no involvement in the QM2 course.} 

\subsubsection{Case study analysis}\label{sec:cases}

For the current analysis, three of the nine interview participants were selected to be case studies. We chose to present case studies because we wanted to provide a detailed, holistic perspective of student engagement with HWCs, and case studies are well-suited for exploratory analyses including multiple sources of data. Further, conducting multiple case studies allows for analyzing multiple perspectives and cross-case themes. Case studies are not intended to be generalizable; however, we intend to provide enough contextual information that readers can make informed decisions about ways in which our results are relevant to other contexts of interest.

Based on the analysis goals, the cases were intentionally chosen to represent variety with regard to engagement with HWCs and the course more broadly. For this selection, we considered the following aspects of the students' engagement in both QM1 and QM2: homework grades, final grades, and number of times doing HWCs in each course. For each of the three students, the following data was used for the case study analysis: final grades (QM1 and QM2), number of times completing HWCs (QM1 and QM2), homework assignments and corrections (QM2), survey responses (QM2), think-aloud interview data (QM2), and responses to general interview questions (QM2). 


In Section \ref{sec:within}, each case study is written up in three parts: case context, think-aloud, and homework assignments and corrections. At the end of each write-up is a summary used to connect the different aspects and present the case coherently. 

The case context section describes the student's performance in QM1 and QM2 and their responses to the surveys and general questions in the interview. This section is intended to give context to the rest of that specific case study. For the open-ended interview questions, we used structural coding to categorize the data \cite{saldana2013coding}. We then used these codes to write a description of each students' performance and high-level perspectives on HWCs.  

\setlength{\tabcolsep}{8pt}
\renewcommand{\arraystretch}{2}
\begin{table*}
\centering
\caption{Framework used to analyze ``Within Structure" HWCs with examples given in all the categories. The comment designated complete and correct adequately identifies an error and what would need to be changed to correctly solve the problem. The comment designated complete but incorrect identifies an error and what would change if the problem had been done correctly (in this case, there were only two quantum numbers, so it is implied that the correct approach required using the other quantum number). However, the quantum numbers all refer to a single electron, so the student's description is incorrect. The comment designated correct but incomplete states the correct answer, but the student does not identify why they initially got the wrong answer. The comment designated incomplete and incorrect does not describe the correct approach (which required considering the symmetry of various spatial and spin components of a two-particle wavefunction) and misidentifies the error (the student's solution was, in fact, normalized, and the normalization was not relevant to why their answer was incorrect).}
\begin{tabular}{cp{6.5cm}p{6.5cm}}
\hline
\hline
& Correct & Incorrect \\
 \hline
{Complete}&
``After revisiting McIntyre (pg 238), I realized that, not only does my integral only consider theta for a single value of phi, it omits the sin(theta) term from the Jacobian. As is, the integral is effectively only viable on a one-dimensional space (basically just $\langle x\rangle$ with a different variable). In order to be valid for a sphere, it needs to evaluate all values of phi (0 - 2pi) and also include the spherical Jacobian." 
& ``I was confused about which quantum number Lz actually acted on, so I assumed it acted on the electron as we previously considered over the hydrogen atom. This is why my answer has a sign difference." \\
\hline
{Incomplete} & ``It turns out that $\lvert E\rvert$ $\sim$ $10^{11}>10^{7}$ by 4 orders of magnitude. Therefore, the stark effect is just a small perturbation of the system." 
&``Wrong, should've written a normalized wave function like the solutions. Since we want this antisymmetry." \\
\hline
\hline
\end{tabular}
\label{tab:coding1}
\end{table*}

\begin{table*}[]
\centering
\caption{Framework used to analyze ``Outside Structure" HWCs with examples given of all the categories}
\begin{tabular}{l|p{10cm}}
\hline \hline 
Additional justification/reflective comment & ``I attended office hours and [Instructor A] explained to me why we can take this. Since, we're integrating over the entire hydrogen atom then we can take the surface area of the entire spherical shell." \\
Small error/typo & ``These should have a factor of $V_0$ but I always forget about coefficients in front of the matrix." \\
Logistical comment & ``This is my final answer for related both expressions. I forgot to highlight and box." \\
Question to grader & ``I realize that the overall complex phase of $\ket{1}$ did not necessarily have to be purely complex or purely real. Given that my answer complies with all the constraints of the problem, is it still valid?"\\
Noted error in solutions & ``The answer key is off by a factor of 1/3 when calculating this. It includes the factor of 1/3 in the algebraic expression, but omits it in the next step once numbers are plugged in."\\
Forgot to address part of question & ``I think I forgot to actually add in the energy value for this state, since it looks like I got too in-depth about the different states which give the same energy."\\\hline \hline
\end{tabular}
\label{tab:coding2}
\end{table*}

Next, we analyzed the students' HWCs. All of their homework submissions, including annotations, were downloaded from Canvas. The first author conducted a round of initial (open) coding \cite{saldana2013coding}, focusing the analysis on student annotations and the corresponding work. After review and discussion, we developed a framework in which comments were first identified as either within the HWCs structure, as established by the instructor, or outside that structure. This framework was used so student engagement could be related to the initial intention when implementing HWCs while still maintaining an exploratory approach. If the comment was identified as within the HWCs structure, it was then analyzed along two axes: ``completeness" and ``correctness." For a HWC to be considered complete, it had to include both a reflection about what was wrong with the initial attempt and a statement about what the correct approach would have been (following the guidance of the instructor). If the HWC was missing one of these pieces, it was identified as incomplete. For a HWC to be considered correct, the student had to have correctly identified their error and what the correct approach would have been. In the case of incomplete HWCs, only the included elements were evaluated for correctness (i.e., if what the student wrote was correct, an incomplete HWC could still be correct). To make these designations, the first author referenced the original homework assignments, the provided solutions, and the textbook. Examples of comments in each of the four categories are shown in Table \ref{tab:coding1}. If a comment was determined to not be within the HWCs structure, it was given one or more subcodes, shown in Table \ref{tab:coding2}; these subcodes were developed through a round of focused coding, based on the initial codes, and revised as necessary. Since students often wrote multi-part comments, none of these codes were exclusive, and it was not uncommon for a comment to be assigned multiple codes. Throughout this coding process, the first author wrote analytic memos \cite{saldana2013coding} to record patterns present for each individual student, as well as emerging themes between the students. 

After conducting our initial analysis of the HWCs, we moved on to the think-aloud portions of the interviews. Using the screen recordings and transcripts, the first author wrote detailed descriptions of each the interviews. Due to the nature of the setting, students were often referencing material on their screen, reading fragments of text aloud, and writing without verbalizing, so this served to create coherent documents to accompany the transcripts to facilitate analysis. During this process, the first author once again used analytic memoing to identify potential trends within and between cases. Then, the think-aloud interviews were coded using a combination of \emph{a priori} codes, based on the trends identified in the students' HWCs, and emergent codes. The focus of this portion of the analysis was to identify ways in which the data from the HWCs and think-aloud interviews supported, clarified, and/or contradicted one another for each student.

We note that due to the interconnectedness of the data sources and the nature of cross-case comparison, the different data sources were coded separately but not independently. We used our codes and analytics memos to guide the case study write-ups; however, we do not report specific codes and frequencies. Frequencies are not meaningful to report because students varied significantly in how they split up their comments, which was reflected in the number of codes.

For the cross-case analysis, we used the individual analyses and analytic memos to identify emergent themes, focusing on notable similarities and differences between the three students. The emergent themes were discussed and revised as necessary. Throughout this process, we revisited the within-case analyses; as an example of why this iteration was necessary, there were instances of all three students spontaneously (i.e., unprompted by the interviewer) bringing up the same topic. While one comment may not have otherwise been included in the individual case descriptions, the consistency across cases motivated its inclusion.


\section{Results}\label{sec:results}

This section will report results in three different subsections. First, we provide descriptive statistics present in students' engagement with HWCs across all 12 courses in our data set. Next, we present detailed descriptions of each of the three individual case studies. Finally, we synthesize the three case studies to identify key similarities and differences. 

\subsection{Descriptive statistics}\label{sec:stats}

In this section, we will report statistics related to the number of students doing HWCs in different courses. A student is considered to have done HWCs if they made any comment or annotation beside one noting they forgot to upload a portion of their assignment or list their collaborators (i.e., no judgment of quality was made). When interpreting these results, it is important to note that in Instructor A's QM2 course (Fall 2023), students were reminded of HWCs during the semester through surveys and an interview solicitation; these interventions did not occur in any other course. This course showed high participation with HWCs but did not show the highest rate of participation in our dataset.

\begin{figure}[h]
\includegraphics[width=0.48\textwidth]{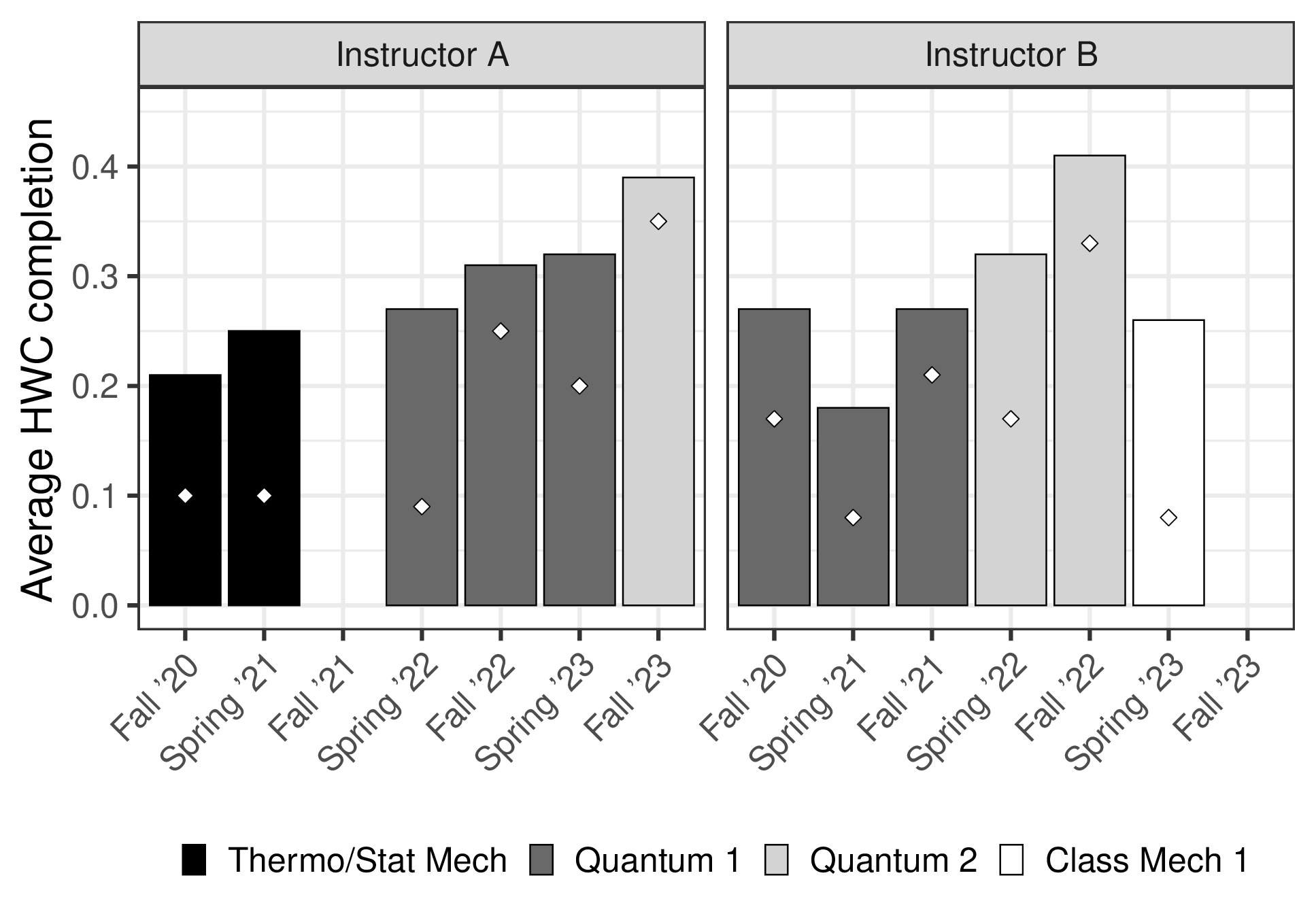}
\caption{\label{fig:courses} Mean proportion of HWC completion (i.e., the percentage of students who did HWCs each week, averaged over all weeks) for 12 classes, split by instructor. The $\diamond$ represents the median proportion of HWC completion in the distribution of students for each course.}
\end{figure}

In Fig. \ref{fig:courses}, we plot the percentage of students who did HWCs each week averaged over the whole semester for 12 different classes, separated by instructor. The mean HWC participation for each course ranges from 18\% to 41\%, with the lowest value for a single week being 11\% and the highest being 55\%. The median participation rate is lower than the mean for all the courses and notably lower for many of them, indicating that a small percentage of students are doing HWCs frequently while most students are rarely doing them. 

\begin{figure}[t]
\includegraphics[width=0.5\textwidth]{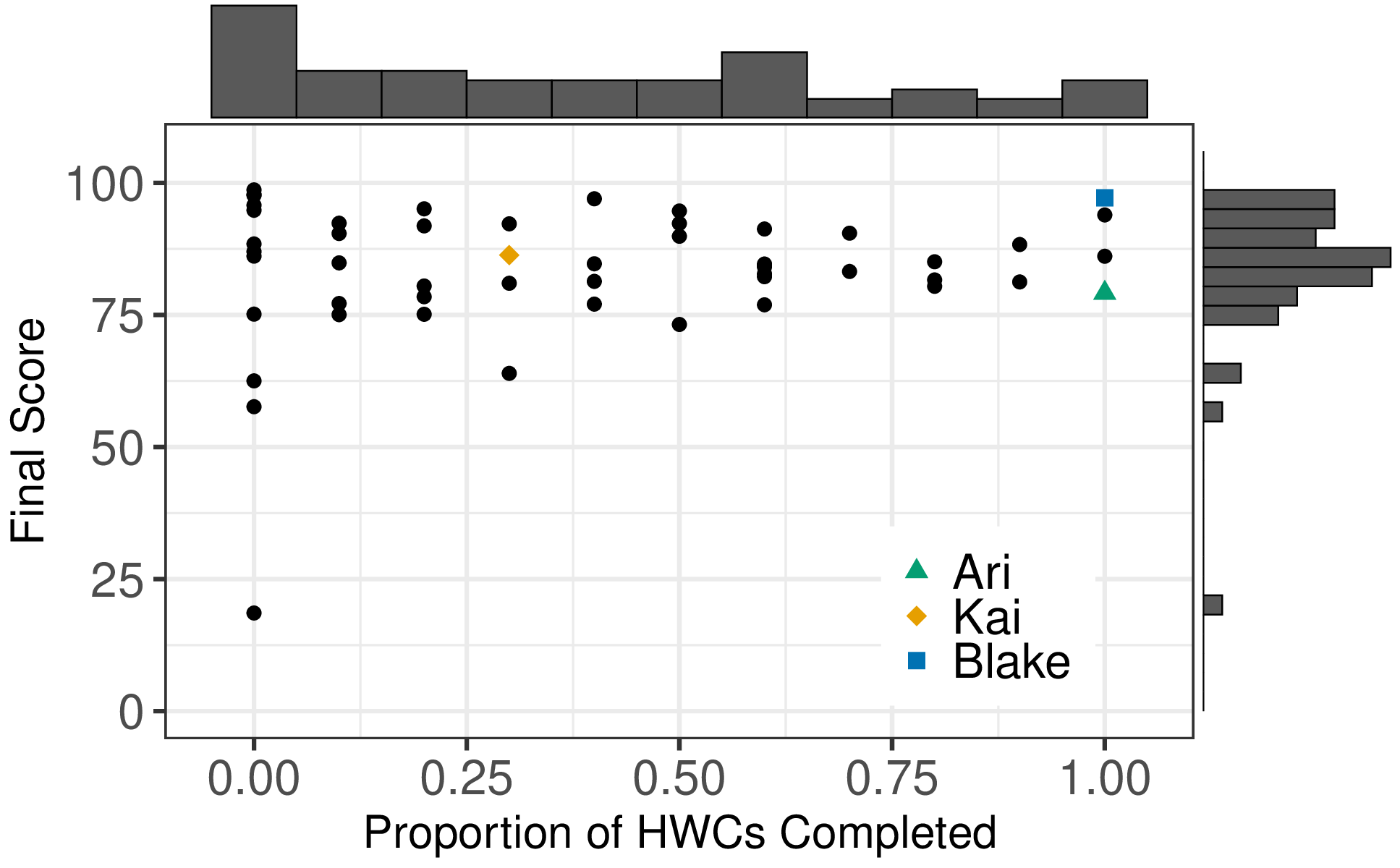}
\caption{\label{fig:IAplot} Students' final scores versus the percentage of weeks they did HWCs for Instructor A's QM2 course, with each dot representing one student. The marginal distributions for each variable are shown on the top and right. The three case study students are identified.}
\end{figure}

\begin{figure}[b]
\includegraphics[width=0.5\textwidth]{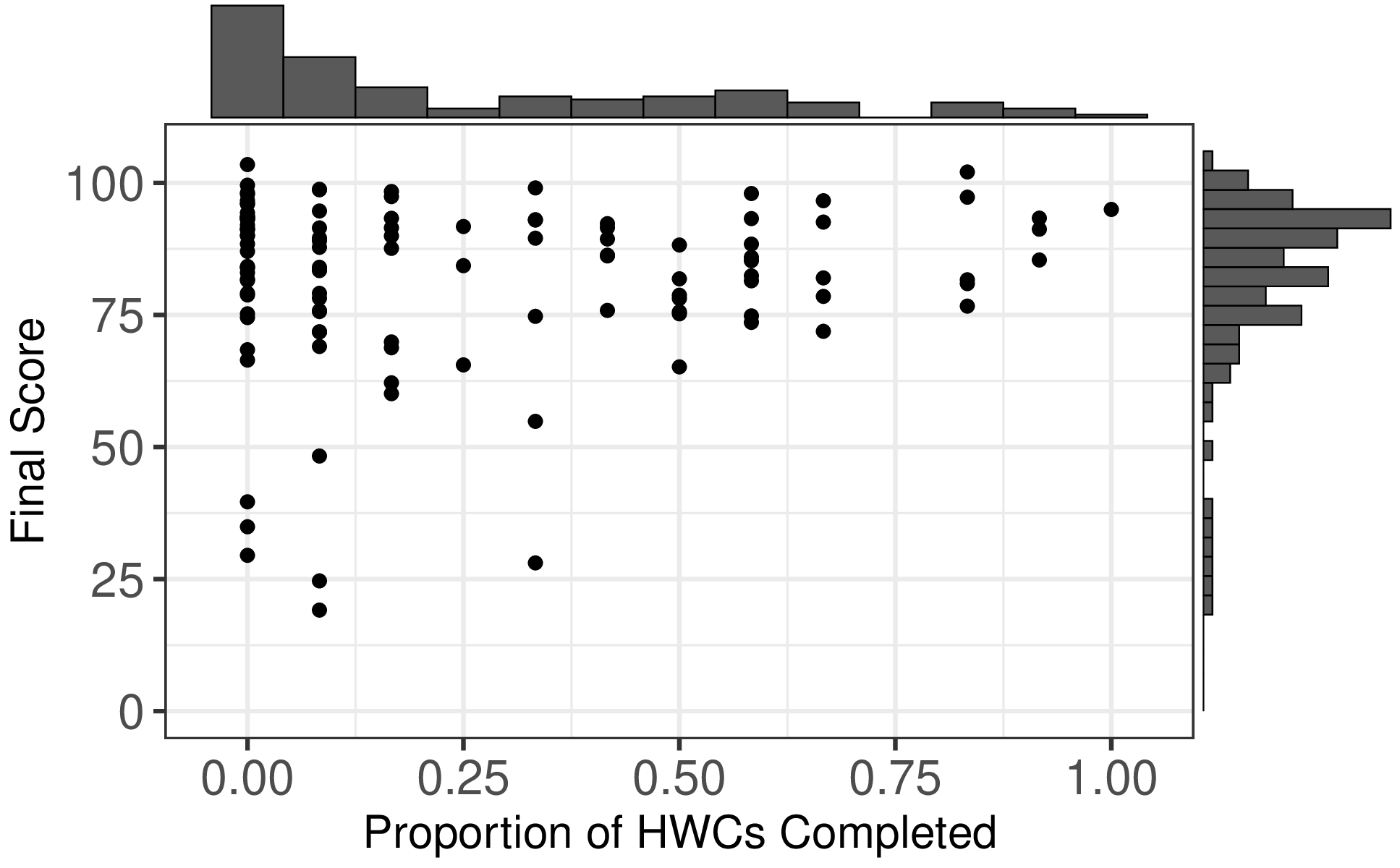}
\caption{\label{fig:IBplot} Students' final scores versus the percentage of weeks they did HWCs for Instructor B's CM1 course, with each dot representing one student. The marginal distributions for each variable are shown on the top and right.}
\end{figure}

In Fig. \ref{fig:IAplot} and Fig. \ref{fig:IBplot}, we plot students' final scores versus the percentage of weeks they did HWCs for Instructor A's QM2 (Fall 2023) course and Instructor B's CM1 (Spring 2023) course.\footnote{The CM1 course was co-taught with another instructor but used the same method of HWCs implementation.} Corresponding plots were generated for all 12 courses; however, only these two plots are included to provide examples that are representative of the larger data set (determined qualitatively). In particular, Fig. \ref{fig:IAplot} represents a very typical distribution and Fig. \ref{fig:IBplot} shows the strongest (though still fairly weak) correlation. The marginal distributions for HWC completion and final score are shown on the top and right of the Fig. \ref{fig:IAplot} and Fig. \ref{fig:IBplot}. The distribution of HWC completion supports the conclusions drawn from Fig. \ref{fig:courses}, showing a positive skew for both courses, with a plurality of students never completing HWCs. The distributions for the other 10 courses qualitatively fall in between the two presented here (again, consistent with Fig. \ref{fig:courses}). 

To address the question of \emph{who} does HWCs, we focus our attention on the scatterplots in Fig. \ref{fig:IAplot} and Fig.
\ref{fig:IBplot}. One might have hypothesized that primarily higher-scoring\footnote{We use higher and lower scoring specifically in reference to students' final grades.} students would engage in HWCs because the same factors driving their high score would motivate engagement with HWCs. Alternately, another possibility would be that primarily lower-scoring students engage in HWCs because they are concerned about their grades or recognize they need additional support to understand the content \cite{griston_homework_2024}. Figures \ref{fig:IAplot} and \ref{fig:IBplot} provide insight into if HWC engagement is dominated by either group of students. Figure \ref{fig:IAplot} shows a fairly even distribution: we see both students who received high scores and comparatively lower scores engaging in HWCs; likewise, we see a wide range of students (with regard to final score) not participating in HWCs at all. In Fig. \ref{fig:IBplot}, we see a slightly more triangular shape. In this course, more students received a lower final grade, and none of those students frequently completed HWCs. This result is not surprising, given that CM1 is typically the first upper-division physics course and required by all majors, while QM2 is typically taken in students' last year and only required for certain degree tracks. This upper-triangular shape also appears in other courses. However, the included plot was chosen because it represents the most significant display of this effect (i.e., this trend was not \emph{more} present for any other course). 

When interpreting the relationship (or lack thereof) between final grade and engagement with HWCs, we note that both instructors explicitly designed their courses with the intention that students who actively participate and put effort into the course should succeed; in other words, students who are scoring significantly below average are likely not engaging in core aspects of the course (e.g., not turning in homework assignments). Therefore, while the experience of these students is important, their lack of engagement with HWCs is likely not specific to HWCs. When focusing our attention on the rest of the students, we still see a broad range of students frequently completing HWCs, as well as a broad range rarely doing so. Hence, in our dataset, HWC correction engagement is not limited to only higher-scoring or only lower-scoring students.

One caveat of the above analysis is that there could exist a category of students who did engage with the review process of HWCs but did not identify any mistakes and so did not leave any comments (and based on our categorization, would then not be labeled as having done HWCs). While it is certainly possible that such students exist, the rest of our analysis leads us to believe that this is not a frequent occurrence. 

\subsection{Within-case analyses}\label{sec:within}

In this section, we introduce three students and discuss how they each engaged with HWCs, with the intention of providing a holistic perspective of each student. For each student, we will first provide contextual information pulled from the full suite of data from that student. Then, we will discuss their engagement with HWCs on a specific homework assignment, using data from the think-aloud section of the interview. Finally, we will discuss trends in their HWCs across the entire semester. 

\subsubsection{Ari}
\textbf{Case context:} Ari received a final grade of a B- in both QM1 and QM2. He completed HWCs on every homework assignment in both semesters. Broadly, Ari expresses a very positive view of HWCs and repeatedly states that they help him understand where he went wrong. Further, he emphasizes that the opportunity to do HWCs reduces his stress when initially completing the assignment. In the interview, he states, “It's good knowing that I can do homework corrections in the end, just so I'm not chasing, like, a perfect score on a homework assignment.” He compares this to classes where HWCs are not offered, saying, “So if I fuck up the homework assignment, I fucked it up. And there's nothing I can do about it. So I'm not learning anything. I'm just chasing a grade now. So I'm not really benefiting my learning experience.” In such courses, he also states that he “only look[s] at the posted solutions when [he] has to start studying.” Moving beyond his own experience, he hypothesizes that without the opportunity to do HWCs, students may be motivated to use external resources, as they are “chasing that high grade.” 

While he is clear about finding HWCs valuable, Ari also discusses some elements of HWCs he has had issues with before. He notes that he sometimes has difficulty understanding the solutions, which limits his understanding of the content. He also notes having inconsistent graders, with some giving full points back and others giving ``a bone to chew on, no full credit." He goes on to explain that he rarely looks at graders' comments and does not ``take their comments too seriously" because ``there's no standard that the graders go by."

Ari stated that he has also been in other courses that offered HWCs. When asked about his instructors' motivations for implementing HWCs, he responds: ``Depends on the instructor. Some instructors probably do it because the department is doing it, so they're like alright, well I've gotta drink the Kool Aid here. Or there are some other instructors that are like, you know what, that is actually very beneficial to the students, I can see it like helping them and stuff like that." When asked about implementation in other courses, he notes that all of them focused on reflection about ``what went wrong" and ``how you went wrong on your thinking."

\textbf{Think-aloud:} In the think-aloud interview, Ari checks his answers with the final answers in the solutions, as well as checking some intermediate steps and equations. He communicates looking for specific expressions, for instance, ``We're looking for this one over j plus one half." In several places, Ari notices discrepancies between his own work and the solutions. He handles such discrepancies in a variety of ways.

On one question, Ari notices a way in which the solutions differ from his work and decides this means he did the question wrong. On this question, the students were asked to confirm several energy level splitting values. The solutions showed a calculation of the first value and then determined the following values by noting how they were related to the first value (e.g., 16 times smaller). Ari, on the other hand, independently calculated all the values without noting how they were related to one another. While comparing his work to the solutions, he states that he got all the correct values but ``was not getting the 16 times smaller or 5 times smaller" and that he will have to correct this question. However, Ari's initial work was correct and this difference was merely a different algebraic approach. 

In other cases, Ari is unable to match the steps between his own work and the solutions but still determines that his answer is right. On one question, in which students were instructed to derive a provided equation, he is unable to find the expression he is looking for, but says, ``I definitely did do this correctly...I got the equations in the book, so I feel good on this question." However, he still leaves a comment reading, ``The results I got there are the same as in the book. The solutions are bit different than mine :(. so I feel confident in this." In some cases, Ari similarly notices differences between his own work and the solutions but decides not to comment on it. In one case, he notices that his work was longer than that of the solutions and states, ``I did do this one right...but I could have done [it] a lot quicker." While Ari does not specify exactly how he decides whether or not to write a comment, he does mention considering previous difficulties with graders. When discussing the solution to one question, he states that he “feels good on this question” but that he will “still leave a comment just because that safety net and prior graders have really ruined [him].”


While the previous examples were all cases in which Ari was correct, there were also cases in which Ari's answer was actually incorrect, and he struggled to identify his mistake. For instance, the first part of one question asked students to draw an energy level diagram and the second part asked them to calculate certain values corresponding to properties of the graph. Ari identified that both his energy level diagram and his calculations were incorrect. He then attributed his errors in the second part of the question to his errors in the first part, saying ``...I messed up the numbers. This is probably because my graph is incorrect." However, he does not pinpoint \emph{how} the error in his diagram led to this secondary mistake. In fact, the second part of the question was referencing a part of the diagram that he had done correctly. 

\textbf{Homework assignments and corrections: } Completing HWCs every week, Ari left a total of 158 comments, averaging about 14 per assignment (compared to an average of 4 and 5 comments per assignment from the other two case study subjects). However, only approximately half of these comments were coded as within the HWCs framework. Many of his comments were providing additional context or explanation, rather than addressing something that was wrong. An example of this type of comment is, “I've used techniques learned in PDEs here. Essentially, solving the Laplacian at the boundary of a circle.” Often, these comments seemed prompted by what was written in the solutions, noting things that the solutions noted, even if they were not mentioned in the homework assignment. 

In many places, Ari reflected on the fact that he had initially been confused about a certain topic or question. In some cases, he commented that even after looking at the solutions, he was still confused. For example, in one comment he writes, ``Even after attending office hours, this proportionality relation is still unclear to me. However, relating the fact that we know that the eigenvalues of the SHO have a form of the solution then this solution makes a bit more sense but not quite." Here, we note that he is not writing this comment to earn points back but is reflecting on the state of his own understanding and communicating his confusion to the grader. 

When within the HWCs framework, some of Ari's corrections were considered correct and complete, providing adequate discussion of his mistake and the correct answer. However, he also wrote corrections that were not entirely complete and/or correct.  In some cases, Ari would simply state what he did wrong without any comment about what the correct answer would be; in other cases, he stated that he ``understood" what he did wrong, without specifically identifying his error. As an example of the latter, in one comment he writes, ``after reading the explanation on the answer key, then it makes sense that each perturbation will be negative in this case." Here, he notes what the correct answer is, implying that his differing answer is wrong, but does not explain \emph{why} he got the wrong answer. Of the comments that were deemed incorrect, some were instances of Ari misidentifying his mistake, as was the case in the think-aloud interview. In other cases, the physics content of his comment was not correct. 

Another pattern present in Ari's corrections (both within and outside the HWCs structure) is that he often made statements quantifying \emph{how} incorrect he was, using phrases such as ``huge oversight," ``hugely mistaken," and ``seriously incorrect." Adding to the tone of his comments, he often included ``sad faces" in his comments, with the character pairing :( appearing 33 times over the course of the semester. 

\textbf{Summary:} Overall, Ari expresses motivation to complete HWCs. He believes HWCs both contribute to his learning and make completing homework assignments less stressful. However, he also notes that when HWCs are not offered in a course, he does not go through a similar process of reflection independently. He writes many comments that are not within the HWCs framework, including reflections on his work and additional information prompted by the solutions; he commonly reflects on his level of understanding of the content, communicating when he is still confused about a topic. In his comments, he frequently uses language implying an emotion or judgment related to what he is writing. While he makes many reflective comments on his assignments, he sometimes struggles to produce HWCs that are both correct and complete. In many cases he writes that he understands his error after looking at the solutions, but his corrections sometimes lack a complete explanation of his issue.

\subsubsection{Kai}

\textbf{Case context}: Kai received a final grade of an A in  QM1 and a B in QM2. He completed HWCs four times in QM1 and three times in QM2. He described finding HWCs valuable but being primarily motivated by improving his grade. When asked about his motivation for completing HWCs, he explained that he would ``look at the week and...go, `how many questions was I not confident on on that homework,'" clarifying that it was ``more of a points thing for [him]." He adds that doing HWCs ``strengthens learning by going over with a fine tooth comb" and specified that he was ``able to figure out some things that helped [his] learning." He also stated that when given the opportunity to do HWCs, he would do his best on each problem on the initial assignment but ``probably not spend so much time on each part of each problem."

When discussing his experience in courses that do not offer HWCs, he said he ``look[s] at the posted solutions for those courses just for studying [for] exams," clarifying that means he is reviewing the solutions ``maybe twice a semester." He goes on to mention that the graders ``don't leave very many comments, or very good comments."

When asked what he thought his instructor's motivation for implementing HWCs was, he said, ``I think the main thing is to get us to go back through the solutions in the first place. And to compare our answer to the solutions to make sure we got the problems right but also to look at...maybe there are different ways of doing the same problem." While most of his comments about HWCs were positive, he did note that sometimes the provided solutions did not have enough information and that good solutions are necessary to ``be able to fully get everything out of the HWCs."

\textbf{Think-aloud}: During the think-aloud interview, Kai reviews the steps in each solution, including for questions that he has already determined he got correct. He notes that his specific strategy depends on the type of question and how confident he felt about his solution. In some cases, he starts by reviewing his own work; in others, he begins by independently trying to understand the solution. He emphasizes that ``the most important part for [him] is [that he] understand[s] how the teacher did it" and that if his answer is right, he will usually only skim through his own work. 

In several cases, he notes that he is trying to identify any differences between his work and the solutions. For one question where he got the same answer as the solution, he asks himself, ``Did I do something wrong? Was it like a fluke that I got that answer? Or did I just do it a little bit differently?" He eventually decides that he just did the problem differently and goes on to proclaim that the method used in the solutions is ``a lot easier" and he ``thinks [he] learned something" by working through the steps of the solution. Similarly, he commented on the solutions to a different question saying, ``That's very interesting, I would not have thought of doing it like that." On another question, he identifies that his method was different than that shown in the solution but decides that his method was ``much easier." For all of these questions, Kai does not leave any written comments. 

For problems on which he does write a comment, he verbalizes struggling slightly to figure out what to write, saying, ``I don't want to sound just like the solutions, but she put it so nicely." However, he emphasizes that he is ``trying to not just write what the solutions said, [he is] trying to show [his] own understanding."

\begin{figure*}[h]
\includegraphics[width=\textwidth]{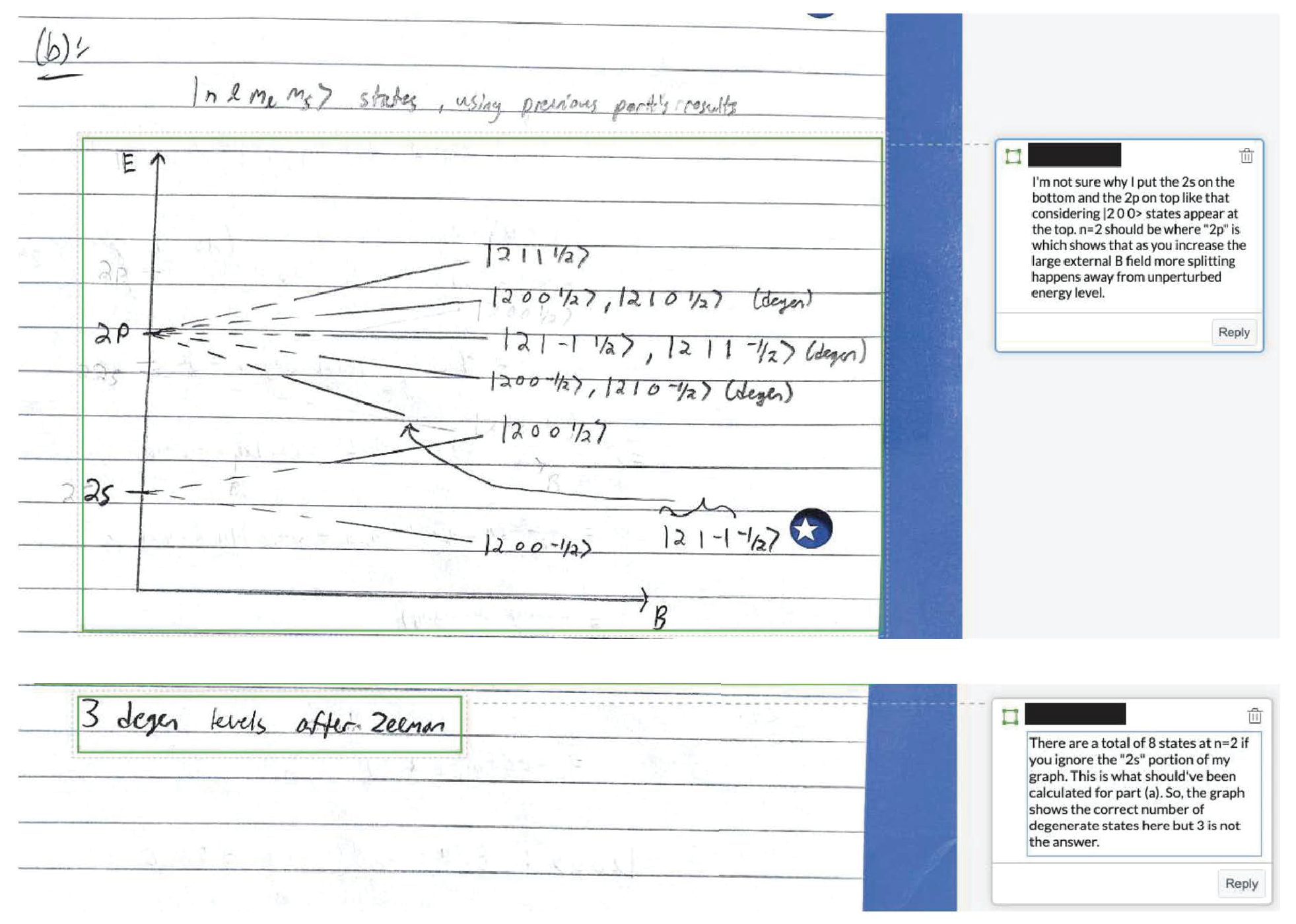}
\caption{\label{fig:S5HWC} Kai's HWCs from the week of the think aloud interview. The question asked the students to show the energy splittings due to the Zeeman effect for the n=2 states of the Hydrogen atom and note any remaining degeneracies.}
\end{figure*}

For several questions, Kai initially struggles to understand the solution. In these cases Kai talks through the steps of the solution, verbalizing the steps or concepts he understands and those he does not. For example, one questions asks the student to draw a graph of the energy splitting due to the Zeeman effect. When reviewing this question, he says, ``I think what I'm not understanding correctly is that here it says n equals two, right, because that's one energy level, right, energy is on the y-axis. So if that's n equals two, there should be eight degenerate states including spin. So as you increase the B-field, these start splitting due to the Zeeman effect, right? So this really should just be n equals two, like the solution, here at the top." Through this verbalization, he decides that he understands the graph drawn in the solution and how his answer was incorrect. However, he still struggled to identify \emph{how} he had gotten his initial incorrect answer and eventually moves on after writing the comments shown in Fig. \ref{fig:S5HWC}. 

At several points throughout the HWCs process, he states that he may return to a problem and work through the steps in greater detail. It was unclear, however, when he was planning on returning to these problems.  He also mentions noting a specific confusion and wanting to ask the instructor about it during office hours. 

\textbf{Homework assignments and corrections}: Kai only wrote a total of 14 comments over the course of the semester. Most of his comments were considered to be within the HWCs framework; however, there were a few instances in which he wrote additional information or a reflective comment or made a minor edit. For his comments within the HWCs framework, several of them were deemed incomplete because they were missing a reflection about what he did wrong. For example, in one HWC, he writes, ``It turns out that $\lvert E\rvert \sim 10^{11} > 10^{7}$ by 4 orders of magnitude..." Here, he states the correct answer, using the language ``it turns out," without addressing why he initially got an incorrect answer. 

\textbf{Summary:} Overall, Kai communicated being primarily motivated by improving his grade but also noted that he found HWCs valuable; however, he only completed them a few times each semester. While he expressed being motivated by earning points back, he spent a significant portion of the think-aloud interview trying to understand the steps of the solution and seeing if he could learn anything from the solution, even for questions that he had gotten correct. He also spent a notable amount of time talking through his understanding of the physics concepts, beyond a line-by-line comparison of the solutions and his own work. In many cases, however, he did not write any comments. When he did HWCs in other weeks, his comments were often short and some were missing a reflection about his initial attempt at the problem. 

\subsubsection{Blake}
\textbf{Case context}: Blake received a final grade of an A in both QM1 and QM2. He completed HWCs six times in QM1 and 10 times in QM2. He discusses being motivated both by improving his grade and better understanding the content. When surveyed, he said his motivation was ``largely just getting points back," specifying that doing HWCs ``was just kind of academically the best thing to do." However, he also notes that HWCs are helpful because he is often ``hovering between I think I get it, and I don't think I get it" and that HWCs help with that. 

Blake also notes that sometimes he did not want to do HWCs because he knew he did poorly on the assignment, describing that ``a lot of times...it'd be like, I know I bombed this assignment, I don't have the wherewithal to go look at all the mistakes I made today." However, he again discusses the motivation of improving his grade, stating that this feeling has not prevented him from doing HWCs as much recently because ``100\% point forgiveness is huge."

While Blake said that HWCs were helpful in that they ``forced [him] to compare [his] answers against the key," he separately reported that he didn't find HWCs particularly valuable. He describes, ``If I didn't understand the problem really well, I didn't understand the problem really well after looking at the answers." He goes on to explain that really understanding the problem would require time that he could not afford to spend. 

\textbf{Think-aloud:} During the think-aloud interview, Blake often references making decisions based on what will cost or earn him points. For example, on one derivation, he had a difficult time deciding if got the same answer as the solutions. He decides that he is okay moving on without writing a comment because he felt ``justified" in his approach. Elaborating, he says he often thinks about what he would do if he were grading the assignment and whether or not he would take points off and that this is often his mentality on the ``derive or show kind of stuff." He states that for this problem, ``if somebody want[ed] to take points away from [him], [he] would probably have a problem."

In other cases, he decides to leave comments in places where he differs from the solutions, even if he does not think his answer was incorrect. For one question, he notes that the question and the answer key are ``a little bit ambiguous as to like what kind of thing explicitly they're wanting." In these cases, he says, he will normally ``leave a correction just in case." Generalizing about this situation, he says he just has to ``wave [his] hands at it and hope for the best." In another instance, he says that he will write a comment to ``feel more justified" and ``cover bases and all that." He also states that he will usually leave a comment if he provides more information than the answer key and that he tries to ``comply with what the answer key has." 

On one question, he struggles to decipher the solutions and decide if/where he made a mistake. The question asked students to make a graph showing all n=2 states; however, he says that the graph in the solutions looks like it is ``just for the 2p states because [the states] all have l=1." He ends up unable to resolve this discrepancy and leaves a comment saying he is not sure if he missed a subtlety in the question explaining why he should have only included 2p states. In fact, the solutions \emph{do} show both the 2p and 2s states. The degeneracies in the system mean the 2s states were not represented by additional lines on the energy graph (i.e., one line is labeled as representing both a 2s and 2p state); additionally, there was a typo in one of the states, making this less clear. Regardless, despite noticing differences between his own work and the solutions, Blake is unable to satisfactorily identify \emph{why} these differences are present. 


\textbf{Homework assignments and corrections:} Many of Blake's comments were considered to be outside the HWCs framework; he frequently added additional context to his initially provided answers. This additional context often seemed prompted by what was written in the solutions. For example, one of his comments on a question he answered correctly reads, ``The square of this term is equal to what the answer key has. I wasn't sure how to plot this, so I chose to manipulate the equation as I did afterwards." 

Blake also uses HWCs as an opportunity to ask a question to the grader. For example, he asks questions about appropriate notation and convention for writing states. He also asks if an answer he provided was valid despite the ways in which it differed from the solutions. 

In a few places, Blake notes times that he was confused when initially doing the homework, as well as times when he was unsure after looking at the solutions. For instance, in one comment, he writes, ``This section has been very difficult for me to understand, and I didn't grasp initially that different in spin can contribute to degeneracy." Here, he reflects on his initial understanding of the content and explains his confusion that led to his incorrect answer. In another comment, he states, ``Also, the answer key gives this as five-fold degenerate, but I have the same states that the key has, and I don't understand where the extra degeneracy comes from" (having written that the system was four-fold degenerate). In this case, Blake is left confused after reviewing the solutions. While at first glance his comment makes it seem as though he simply misread the solutions, his other work makes it apparent that his misunderstanding arose from not understanding that a wavefunction for two particles at different energy levels can be written either as a symmetric or antisymmetric (spatial) wavefunction. 

In a somewhat similar case, Blake wrote a comment reading, ``I think I must have either entered the wrong values into Mathematica or transcribed them incorrectly, because I can't spot where my integrals differed from what is on the answer key." Here, Blake attributes his incorrect answer to a computation error; however, his work demonstrates that he actually set up in the integral incorrectly. 

While some of Blake's comments within the HWCs framework were considered incorrect, others were deemed both complete and correct. For example, one question asked students to identify the eigenstates, energy, and degeneracy corresponding to the second excited state of a 2-D harmonic oscillator. On his initial submission, Blake answered that $E=5\hbar \omega$, corresponding to the nondegenerate eigenstate $\ket{n_x = 2, n_y = 2}$. For his HWCs, he wrote the comment, ``I realize now that this is actually the fourth excited state, since the energy corresponding to the 2-d oscillator is the sum of both 1-d energies. I think what got me confused is that the in-class tutorial we did on this considered the infinite square well, where n=0 is not allowed, and the second excited state is $\ket{1,1}$." This was followed by, ``For the Quantum SHO and not the infinite square well, the second excited state has $n_x + n_y$ = 2, and thus is three-fold degenerate since $\ket{0,2}$, $\ket{2,0}$, $\ket{1,1}$ have the same energy." Here, we can see the clear steps of Blake identifying his error, reflecting on the confusion that led to his error, and noting what the correct answer is. It is  worth noting that while Blake addressed that he wrote the wrong state, he did not explicitly address that he was also incorrect in declaring this state as nondegenerate; however, this was still deemed to be a complete HWC because he addressed this aspect of the solution when discussing the correct answer. 

Over the course of the semester, there were also three different times that Blake wrote a comment accurately identifying an error in the solutions. In two of those cases, the error affected the final answer and Blake was arguing that his differing answer was actually correct. In one case, he was pointing out an error that did not actually impact the answer (a factor that ended up being multiplied by zero). 

\textbf{Summary:} Overall, Blake presents as a high-performing student who is often motivated to do HWCs but provides conflicting reports about whether or not he finds them helpful in understanding the content. He is consistent, however, in noting that he is motivated by improving his grade. During his think-aloud interview, he decides whether or not to make corrections by considering if he thinks the grader will take off points. In his homeworks, we see this reflected in additional comments that seem prompted by what was written in the solutions.

Some of Blake's HWCs were both complete and correct, including appropriate reflection; however, there were also several times when he was either unable to identify his error or misidentified his error. In addition to formal HWCs, we also see Blake writing comments to reflect on his own understanding of the content, ask the grader a clarifying question, or note an error in the solutions.

\subsection{Cross-case analysis}\label{sec:between}

In this section, we discuss the themes present across the three case studies. These themes are not intended to be independent from one another, and we will note connection where necessary.

\subsubsection{Broad and varied use of HWCs structure}

When implemented, HWCs were intended to provide students the opportunity to compare their work to the provided solutions, make corrections, and earn points back. Within this process, we see variety in engagement between the three students. For example, in the think-aloud interview, Ari verbalizes checking for specific steps in the solutions to verify his own work and moves on from questions once he has decided he is confident in his answer. In some cases, he notes that the provided solutions follow a different process, but there are times that he moves on from a question without understanding the steps in the provided solution. Blake, similarly, moves on from questions after deciding he feels confident enough that he will receive full points for his submission, even in cases where he is unable to confirm that his work is equivalent to that shown in the solutions. While both of these students often made an effort to understand the solutions, they did not deem understanding the solutions a necessary step if they were already confident in their work. Kai, on the other hand, communicated prioritizing understanding the solution and would spend time working through the provided solution, even after determining that his answer was correct. Additionally, in the think-aloud interview, Kai verbalized his understanding of relevant physics concepts, seemingly trying to determine if he understood the steps of the problem. In contrast, both Ari and Blake make brief statements regarding relevant physics concepts but do not talk through them as extensively. 

While all three students completed HWCs that fit within the intended framework provided by the instructor, they also all used the structure provided by HWCs for purposes that were not within the defined HWCs framework. For example, all of them made comments either reflecting on their work or providing additional information in places where their initial attempt was not incorrect nor incomplete. Kai and Blake both used these comments to note when questions or topics had initially been confusing or were still confusing after reading the solutions. These two students also asked questions to the grader about remaining confusions. Again, we see variation within this realm of engagement as well. While all three students used the HWCs structure to make comments outside the framework, they left different types of comments from one another and did so with different frequencies.

\subsubsection{Struggle to identify root error}

While all three students made corrections that were sufficiently complete and correct, all of them also struggled at times to identify their errors. Specifically, the students struggled to identify the \emph{cause} of their incorrect answers, which meant they were unable to address these root missteps or misunderstandings. This issue took several forms. In some cases, the student merely implied that their answer was incorrect by making a comment stating the correct answer but never actually explained what they had done wrong. Ari, for example, writes ``after reading the explanation on the answer key, then it makes sense that each perturbation will be negative in this case." This statement is in direct contradiction to his initial answer, implying his initial answer was incorrect; however, he never identifies why he initially got an incorrect answer or what step in the solutions differed from his own logic. 

In other cases, the student identified a potential issue but did not verify what exactly was the cause of their error. For example, when Blake mistakenly attributed his error to a typo in Mathematica, he used the language ``I think I must have..." to describe this proposed error, implying he did not actually check that correctly plugging in the numbers would give him the correct answer (in fact, this would not have given him the correct answer). In situations like this, the students were able to identify a potential cause of their incorrect answer but did not follow through to confirm whether or not what they identified was actually the cause. 

Similarly, there were times when the identified error was too broad to be meaningful. For example, when writing a HWC on a question about Clebsch-Gordan coefficients, Ari stated, ``So, obviously I'm unable to read the tables but it's much clearer to me now!" While it may be the case that he had difficulties reading the table, he does not make any specific statements about what was wrong (e.g., he read the wrong row/column, he did not know which quantum numbers were relevant, etc.). 

Finally, there were several times that the student stated that they were unable to resolve the error. On one question, Kai stated that he was unable to identify the discrepancy between his own work and the solutions; he then left a comment noting that he had written an incorrect multiplicative factor of $\frac{1}{2}$. This factor did not actually affect his final answer because it was multiplied by zero. He verbalized, ``I don't know exactly what to do...like on the homework corrections. But I do know that this is just wrong, because I'm missing a factor of $\frac{1}{2}$. So even though it doesn't matter...I think I'll leave that there." Here, Kai knew that he did not fully address the inconsistencies and chose to leave a comment regarding something he was certain was wrong, even though it was not a significant error. 

In all of these cases, the fact that the student was unable to identify the cause of their incorrect answer meant they could not fully reflect on and address the issue. In some of the cases, the students were aware of this limitation; however, in many of these cases, they did not indicate an awareness that they had not fully addressed and resolved the discrepancies. 




\subsubsection{Dependence on content of solution}

An essential step of implementing HWCs is providing students with worked-out solutions. In the three case studies, we saw notable dependence on the content of the solutions: student behavior and written HWCs depended on specific surface features of the solutions, such as the specific algebraic steps shown and supplemental information provided. Additionally, students spoke of ``complying with the answer key" and ``what the solutions wanted," giving insight into students' perspective on the role of the solutions. 

In several cases, students verbalize comparing their work to the solutions by identifying the presence of specific steps. For example, during the think-aloud interview, Ari compares his work to the solutions by checking surface features and then struggles to resolve discrepancies when the surface features are not the same, even though the difference was merely due to an equivalent algebraic process. In this case, he was still confident that his initial submission was correct; however, there were instances in which he had done the problem correctly but made a comment writing he had done something wrong because his steps differed from those shown in the solutions (an example of this is provided in Section \ref{sec:within}). 

In a similar fashion, Kai paid close attention to specific algebraic steps when working through the steps of the solution. For one question, the students were asked to plug numbers into an equation to confirm the values given in the textbook. Kai noted that he had gotten the correct answer but that there were differences between his work and the solutions and was not sure if he had done the problem incorrectly. Eventually, he decided he had just done the problem slightly differently but that the method used in the solutions was ``a lot easier" because she had written the expression in terms of the fine-structure constant and could then plug in $\frac{1}{137}$ instead of all the individual constants. In this case, he successfully determined that his answer was correct and was able to figure out why the steps were different; however, it is noteworthy that he spent several minutes trying to resolve this issue and deemed the method in the solution ``a lot easier" when the only difference was the method of plugging in physical values (which he had originally done in Mathematica). 

In addition to the focus on specific steps when comparing his work to the solutions, Ari wrote comments providing additional, often unnecessary, information based on what was written in the solutions. For example, one question asked students if a certain expectation value depended on time. Ari gave the correct expression for the expectation value and stated that it did not depend on time. While this was already an answer that would have received full credit, he additionally left a comment writing, ``The coefficients do not depend on time. Despite the sum still being there." It is reasonable to believe that this comment was prompted by the fact that the solutions had explicitly noted that the coefficients were not time-dependent. 

This reliance on the solutions is supported by students' comments explicitly noting the importance of good quality worked-out solutions and the difficulties they have had when the provided solutions were not sufficient. In fact, all three students brought up this topic unprompted. Kai noted that sometimes the solutions do not provide enough information and that ``you need good solutions...to really be able to fully get everything out of the HWCs." Ari reflects that he has previously had difficulty understanding the solutions saying, ``There are some solutions I've seen and I'm like, `I don't know what the fuck they just did.'" Blake mentioned that there have been several times he has noticed errors in the solutions; several of his HWCs reflect this statement. 


\subsubsection{Willingness to express confusion}

In the provided instructions for completing HWCs, students were told to make reflective statements about what they did wrong, but they were not explicitly instructed to reflect on their initial or current level of understanding of the relevant question or topic(s). Nevertheless, all three students made statements reflecting on their level of understanding. Specifically, the students noted their confusion about both specific questions and course topics. All three students made statements, either in the think-aloud interview or their HWCs, reflecting on their initial confusion when completing a problem (e.g., ``I was confused on this part...", ``I remember being confused on this one."). This behavior is not surprising, as identifying an aspect of confusion is closely related to identifying an error. More notable, however, is that both Ari and Blake wrote HWCs in which they noted that they were still confused after completing their corrections. Blake, for example, states ``I'm really confused about this, since my value for a is inverted from what the answer key has, but I still have the proper expression from McIntyre." In this case, he identifies a continued confusion, in that he is unable to reconcile his own work with the provided solution. 

On several occasions, Blake also uses HWCs as an opportunity to ask the grader questions about notation and the validity of his solution, which indirectly communicates a lack of clarity. In one comment, Ari writes, ``Even after attending office hours, this proportionality relation is still unclear to me. However, relating the fact that we know that the eigenvalues of the SHO have a form of the solution then this solution makes a bit more sense but still not quite." Here, he notes how his level of understanding has changed over time and that he still does not fully understand the solution. Realistically, it is possible for students to make corrections and receive full credit without fully understanding all aspects of the problem.\footnote{This type of engagement is certainly not the goal of implementing HWCs; however, this issue is a consequence of our inability to directly measure understanding and is not unique to HWCs.} This type of comment is noteworthy, therefore, because it is not required by the HWCs structure and clearly not motivated by earning points; rather it communicates authentic metacognitive reflection. 

\subsubsection{Changing initial homework process}

All three students communicated that the opportunity to do HWCs impacted their attitude while initially completing the homework assignment. Ari notes that the opportunity to do HWCs makes him feel better because he knows if he messes up, he can ``get retribution." Kai states that when HWCs are available, he still tries his best on each problem but may not spend as much time on each part. Blake says that HWCs make him less stressed about his homework but potentially make him less diligent as well. He specifies that if he is working on a ``really, really tricky problem" and is unable to attend office hours, he will ``maybe wave [his] hands at it and just hope to pick it up in corrections." Generally, we see these students experiencing reduced stress when completing homework; one student expresses a potential lack of diligence, but this is not made out to be a frequent occurrence for any of the students. We also note that the student who expressed this sentiment received one of the highest overall grades in the course. \\

\section{Discussion}\label{sec:discussion}

In our analysis, we identified several noteworthy themes and takeaways regarding engagement with HWCs. Our discussion will first focus on important elements of student engagement, addressing the first four research questions, and then move on to specific instructional implications, addressing the fifth research question. Finally, we will summarize our primary takeaways and directions for future work.

\subsection{Student engagement and alignment with student and instructor motivations}

In Section \ref{sec:stats}, we reported that a fairly low percentage of students are actually completing HWCs. We see a small percentage of students completing HWCs frequently, but most of the students are rarely, if ever, doing HWCs. In our previous work \cite{griston_homework_2024}, some instructors expressed concern about HWCs inflating students' grades; while it is true that the implementation of HWCs can improve students' homework scores, our results may ease these concerns due to the low percentage of engagement. It is worth noting, however, that these results may be different in courses where homework grades are typically lower, as the courses analyzed had high homework averages before implementing HWCs. When surveyed, the plurality of students who had not done HWCs said they had intended to but forgot. It is entirely possible that students may be more motivated to do their HWCs (or remember to their HWCs) if their homework scores are lower. With regard to \emph{who} is doing HWC, we found a fairly even distribution, with both students who performed well and those who performed comparatively poorly in the course overall engaging with HWCs.

Regarding the qualitative aspects of HWCs, we see a wide variety in engagement, both within and between the two courses. In Instructor A's course, we see students using the structure of HWCs for many different purposes. While all three students completed a number of HWCs that were within the intended structure, they also wrote separate reflections, asked questions to the grader, and noted errors in the solutions. While such engagement was not necessarily the intent of implementing HWCs, we would be remiss to ignore these behaviors, and we consider them to be indications of productive reflection. One student, for example, wrote  broader comments reflecting on his understanding of the material, which was not part of the HWCs framework but certainly aligns with the broader goals of its implementation.

We also saw notable variation within the defined HWCs structure, both in terms of student intention and behavior. When asked about their motivation, all three students mentioned both improving their grade and learning from the process; however, they emphasized or prioritized different aspects. It is reasonable to expect that students with different motivations will engage with HWCs differently. For example, a student who states that their primary motivation for completing HWCs is to earn points back on their homework may be less likely to engage with the solutions beyond verifying their initial answer. We did see variation of this type, with Kai specifically emphasizing learning from the solutions and understanding the way the instructor did the problem, even when his answer was correct, while the other two students communicated less of a focus on this aspect. However, these behaviors were not as tied to their stated motivations as had been expected. Kai, for example, stated that his motivation was based on ``points" but engaged in ways that were not related to earning points back. Here, we see that the connection between students' stated motivations and their behavior is complicated, which is an important takeaway when considering the instructional implications. Related to this variation, we also saw evidence of one student engaging in self-explanation while completing HWCs (indicated by his verbalized explanations of the relevant physics concepts while reflecting on a problem he had done incorrectly), while such evidence did not exist for other students. It is important to note that due to the nature of HWCs, evidence of self-explaining was only apparent during the think-aloud interview, and we would not expect to detect it on the HWCs themselves. Considering the demonstrated benefits of self-explaining, it may be beneficial if efforts going forward more explicitly investigate if and when self-explaining is occurring and address how to motivate this behavior from more students when engaging with HWCs.  

We also saw a number of examples of students struggling to identify their errors. Specifically, students had difficulty determining the cause of their mistakes. In some cases, the students were aware of this difficulty; however, in many cases, the student named a potential source of error that was either incorrect or too broad to be meaningful. There were also multiple instances of students providing a satisfactory explanation for the correct answer without actually addressing how they had gotten their initial answer and what had been done incorrectly. In a number of situations, this prevented the student from adequately addressing their misunderstanding or incorrect step. In some of these cases, it is also possible that students' difficulty lay in communicating their error, as opposed to identifying it. These aspects are, of course, related, but it is important that students receive support in both aspects. The potential impact of the structure of HWCs on these difficulties will be discussed in the following section. 

Here, it is also important to note that the goal of self-assessment is not necessarily accuracy of the assessment and that students can benefit from engaging in reflection even when misdiagnosing their mistakes \cite{andrade2019critical}. In other words, we have focused on students' struggle to identify their errors not because we necessarily care \emph{that} they are wrong; rather, it is important and interesting to identify \emph{why} they are wrong and how this relates to the potential impact of HWCs on their learning.

For example, the way in which students interact with the solutions communicates what they view as important elements of a solution and what metrics should be used to decide if an answer is correct, which gives insight into their epistemic beliefs. In several cases, students used surface features (e.g., specific intermediate expressions) to compare their own work to the provided solutions and struggled when the surface features did not match. In fact, there were instances in which students mistakenly decided their own work was incorrect because of these discrepancies. We also saw students writing comments that were not necessary and were seemingly motivated by what was written in the solutions.\footnote{We are not claiming that writing unnecessary comments is necessarily negative; rather, we are commenting on the rationale for doing so.} This behavior shows similarities to the strategy of ``pattern-matching," which students have been shown to engage in when using worked-examples to support problem solving \cite{wu2022use}. While some part of this can be attributed to a student's motivation to improve their grade, i.e., they have implicitly decided that it is most important to ``comply" with the solutions, there are also important takeaways about students' views on the nature of problem solving and their epistemic commitments, specifically related to who or what they view as the source of authority. While extensive work has been dedicated to studying student interaction with instructor-provided worked-examples, these examples are typically assumed to be free of errors; hence, there has been no evaluation of student trustworthiness of the solution. In reality, instructor-provided solutions are often not free of mistakes, particularly in the context of the long and complex problems typical of the upper-division level, so more work is necessary to delve into students' perspectives on the source of authority in such situations. Further work on the trustworthiness of knowledge and the source of authority in the context of problem solving is particularly important given the rapid growth of generative AI, which poses a new and complicated source of information for students, which we can expect to become relevant in the context of homework solutions. 

Turning our attention to affect and error climates, students in Instructor A's course seemed comfortable addressing their errors and communicating continued confusion. All three students, at some point, communicated that HWCs were an opportunity to learn. They also mentioned that the opportunity to do HWCs reduced their stress when initially completing homework assignments. We see students' willingness to engage with their errors, including when there was no grade benefit for doing so, as evidence of a productive error climate, and all three students seem to associate this, at least in part, with the existence of HWCs. It is unclear, however, how much of this impact is a result of student engagement with HWCs versus the implicit message communicated by an instructor who implements HWCs (i.e., it is possible this effect would still be present for students who were in a course that implemented HWCs but never engaged with them). There exists significant work discussing error climates and the factors influencing them; however, most of this work has focused on younger populations, and none of it has focused specifically on college-level physics. Considering the existing culture in physics and the importance of positive error climates, research on error climates in physics specifically would be incredibly valuable. 

In the discussion of student affect, it is important to note that one student expressed struggling to complete HWCs because he did not want to look at the mistakes he had made. While this is an important factor to consider, it is also worth noting that HWCs, as implemented here, are optional: if a student finds the process stressful, they do not have to participate. It is also possible that with continued engagement with HWCs and the normalization of making mistakes, such students will find this concern to be less of a deterrent. Research on student affect, and changes in affect, as they engage with their errors would provide additional insight into this topic. 

\subsection{Instructional implications and considerations}\label{sec:implications}

Here, we present more specific details about factors for instructors to consider when implementing HWCs. While this paper focused on one method of implementation, this work has involved extensive discussion with instructors and consideration for other aspects of implementation to consider.

First, we consider the broad structure of HWCs. For Instructor A's courses, students completed HWCs before receiving any feedback and then their assignments were graded after completing HWCs. This choice was made to avoid additional grading workload. However, for instructors with the capacity to do so, it is also possible to provide students with feedback, have them complete HWCs based on the feedback, and then give them a grade based on their HWCs. In this case, students are receiving more support in identifying their errors but consequentially get less practice with the identification step. 

Instructor A also made the choice to give students full credit for appropriately corrected problems. In part, this choice was also made to reduce the grading work load: to give students partial credit, the grader must separately grade the initial submission and the HWC/reworked solution, but when giving students full credit, a grader does not necessarily need to assign a grade to their initial submission. To reduce the grading workload even further, some instructors have opted to grade HWCs based on completion. When thinking about grading, it is also worth considering the instructions and scaffolding provided for graders. While not part of the analysis in this paper, instructors have noted differences between expectations for graders and actual grading behaviors; this is supported by our data, with students noting inconsistencies between graders. 

In our previous work \cite{griston_homework_2024}, some instructors indicated concern regarding giving students full credit for doing HWCs, as they were worried about inflating homework grades and the potential for students to ``game the system" by putting minimal effort into their initial attempt and then completing HWCs to earn full credit. While these are certainly valid concerns and important for instructors to consider, we note that in our dataset, we saw relatively low participation in HWCs, even with the opportunity to receive full credit for doing so. Additionally, with regard to the possibility of students taking advantage of HWCs, we saw little, if any, evidence of this in our data. Further, we think it is important to consider the alternative: how are these students engaging with homework assignments when there is no opportunity to do HWCs? If a student, when offered the opportunity to do HWCs, chooses to not engage authentically with the homework assignment and then take advantage of HWCs, it seems reasonable that if they were not offered the opportunity to do HWCs, they would potentially seek out other sources of support (e.g., online solutions). 

Within the broader structure, another factor to consider is the timeline of the HWCs process. Instructor A chose to post solutions soon after the initial assignment deadline and then gave students a short period of time to complete HWCs. The benefits of this structure are that students can do HWCs soon after completing the initial assignment, students are not delayed in starting the next week's assignment, and grading can happen more quickly. However, this structure means that instructors cannot accept late work since they post the homework solutions immediately after the original submission deadline (unless they are comfortable accepting late work with the solution already available). Alternatives to this structure include giving students a longer period of time to complete HWCs (and delaying grading), or building HWCs into the following week's assignment. 

With regard to the framework of the HWCs themselves, there are a few factors to consider. Instructor A  required students to make a reflective statement about what they had done wrong but did not ask students to fully rework the question (a choice also made by some other instructors). Variations can be made with respect to the content of the reflection (e.g., specifically requiring a statement about how to avoid the mistake in the future) and the level to which the correct solution needs to be provided. These choices may be influenced by the instructor's goals for the course. For example, an instructor teaching a Math Methods course, in which the goal is to improve students' mathematical fluency when solving physics problems, may find it more important for a student to have correctly worked through all the steps of a problem than an instructor teaching a course where the priority is conceptual understanding.

There are also important implications with regard to the aforementioned concern of students' difficulties with identifying their errors. For example, a student who does not fully rework the question could misattribute an error to a mistake in the previous portion of the problem and move on, theoretically receiving points for the HWC if the grader did not verify this. For a student who had to fully rework the question, however, it is less feasible to misattribute the original mistake to an error in the previous portion. Of course, it is certainly possible for students to correctly rework a problem and still misidentify their mistake; this is merely a point to consider when comparing different methods of implementation.  

Finally, an element of HWCs for instructors to pay attention to is the content of their provided solutions. In our data, we found students interacting with the solutions in ways the instructor had not expected, and it is important to consider that someone who is learning content for the first time will likely interact with a worked-out problem differently than someone who is comfortable with the content. This does not mean that instructors should necessarily provide more detailed, error-free solutions, merely that it is worth being intentional about the content of the solutions. For example, we observed cases in which students were confused about which of the steps in the provided solution were essential to the solution versus just made for algebraic convenience. As such, instructors may want to be considerate of if and how they are differentiating these types of steps in their provided solutions. 

\subsection{Primary takeaways and future work}

In this paper, we have considered the implementation of HWCs, which are intended to motivate students to engage in reflective practices as part of the problem solving process. We see students engaging with HWCs in a wide variety of ways, both within and outside of the intended structure. Based on our results, we believe that HWCs can be a useful pedagogical tool with a number of potential benefits to students; however, we certainly see areas for improvement with regard to implementation.

A primary takeaway from our results is that even upper-division students need additional scaffolding to successfully compare their work to provided solutions, identify meaningful discrepancies, and address their mistakes. While the implementation of HWCs may motivate students to engage in the reflection process, it is not a given that students will do so effectively without additional support. It may be useful for future work to focus on implementing self-diagnosis activities in upper-division courses, focusing on homework assignments rather than quizzes or exams. It may be that this higher level of structure is not necessary throughout the semester and could be faded towards the format of HWCs as discussed in this paper (or something similar).

While the work presented in this paper illuminates important aspects of student engagement in reflection during problem solving, it is limited in scope; specifically, we focused primarily on a single course, instructor, and method of implementation. Further work is necessary to characterize student engagement for different courses, student populations, and methods of implementation. Additionally, our data do not allow us to make any causal statements about the impact of HWCs on learning. Future studies could be designed with HWCs as an intervention to better investigate this potential effect of HWCs. While it is certainly worthwhile to consider the impact of actually doing HWCs, we are particularly interested in the impact of implementing HWCs on classroom error climates and how improvements to error climates may benefit student learning and problem solving skills. 

\section{Acknowledgments}
Thank you to the CU PER Group for providing feedback on this project. This work was supported by funding from the Department of Physics at the University of Colorado Boulder, the NSF GRFP, and NSF Grant No. 2143976. 

\bibliography{HWCs}

\end{document}